# Sequences, Items And Latent Links: Recommendation With Consumed Item Packs

Rachid Guerraoui
EPFL
rachid.guerraoui@epfl.ch

Erwan Le Merrer
Technicolor
erwan.lemerrer@technicolor.com

Rhicheek Patra
EPFL
rhicheek.patra@epfl.ch

Jean-Ronan Vigouroux
Technicolor
jean-ronan.vigouroux@technicolor.com

*Abstract*—Recommenders personalize the web content by typically using collaborative filtering to relate users (or items) based on *explicit* feedback, e.g., ratings. The difficulty of collecting this feedback has recently motivated to consider *implicit* feedback (*e.g.*, item consumption along with the corresponding time).

In this paper, we introduce the notion of *consumed item pack* (CIP) which enables to link users (or items) based on their *implicit* analogous consumption behavior. Our proposal is generic, and we show that it captures three novel implicit recommenders: a user-based (CIP-U), an item-based (CIP-I), and a word embedding-based (DEEPCIP), as well as a state-of-art technique using implicit feedback (FISM). We show that our recommenders handle *incremental* updates incorporating freshly consumed items. We demonstrate that all three recommenders provide a recommendation quality that is competitive with state-of-the-art ones, including one incorporating both explicit and implicit feedback.

## I. INTRODUCTION

In this Zetabyte Era, the abundance of information calls for *personalization* systems to ease the navigation of users. Among these systems, *recommenders* are becoming mainstream, and are used by major service providers such as Facebook, Amazon and Netflix. Some recommenders make use of the content of the items: these include *popularity-based*, *knowledge-based* or *demographic-based* schemes [8]. Others are *content-agnostic*: these are mainly *collaborative filtering* (CF) [14], [44] schemes, and are predominant today for they achieve good recommendation quality without requiring any prior knowledge of the content of the items recommended. Recommenders typically collect user *preferences* using *explicit* feedback [32], such as *numerical* ratings (star ratings in Imdb, Netflix, Amazon), *binary* preferences (likes/dislikes in Youtube), or *unary* preferences (retweets in Twitter).

Yet, relying on explicit feedback raises issues regarding feedback *sparsity* (in systems where the item catalog is large, users tend to give feedback on a trace amount of those items, impacting the quality of recommendations [8]), and limited efficiency for recommending fresh items in reaction to recent user actions [37]. A few implicit recommenders have been proposed to answer those shortcomings. Some leveraging the context [9], [16], [18], and some generating pseudo-ratings using *launch time* and *purchase time* for items [36]. While the former relies on knowledge about the content, the latter does not compete in terms of quality with explicit feedback based schemes, like singular value decomposition (SVD) based algorithms [34].

### A. Motivation and Challenges

The most recent alternatives [10], [17], [33], that show competitive results with SVD recommenders, aim at uncovering high-order relations between consumed items. Each paper proposes a specific algorithm, with an arbitrary definition of sequences of consumed items. Our motivation is to investigate the existence of a higher level abstraction for sequences of consumed items, and algorithms for dealing with them. Such an abstraction, we name a *Consumed Item Pack* (CIP), allows to reason about and to propose sequence-aware algorithms within the same framework, capable of addressing implicit recommendation.

The challenges are threefold. *(i)* We first have to highlight that the notion of CIP captures the analogous consumption pattern of users (*e.g.*, the one exposed in [17]). We demonstrate this in § II, where CIP-based item communities are uncovered on a concrete video consumption dataset. This notion must be generic enough so that different algorithms could be devised using it. *(ii)* The next challenge is computation complexity of the proposed algorithms in the CIP framework. Leveraging CIPs for building implicit recommenders is not immediate for the computation time can easily become prohibitive given the sizes of user consumption logs in production systems. This is, for instance, the case in previously introduced sequential approach HOSLIM [17], where algorithmic tractability is at stake. Concerning memory-based CF, we show in § III-A (resp. § III-B) how to build a CIP-based similarity metric that is *incremental*, which helps in designing an implicit user-based (resp. item-based) recommender that *scales* while providing good recommendation quality. Moreover, we also present a model-based CF technique incorporating CIPs in § III-C which leverages neural word embeddings [40]. We demonstrate that our techniques scale with an increasing number of computing nodes while achieving a speedup comparable to Spark's Alternating Least Squares (ALS) recommender from MLLIB[1] library. *(iii)* These proposed implicit algorithms have to provide accuracy that is at least comparable with classical CF recommenders, in order to be adopted in practice. For assessing their performance, we then conduct a comparison with an explicit SVD-based recommender [34], with an implicit one [36], as well as with a recent state-of-the-art algorithm [39] incorporating both implicit and explicit techniques

---
[1] http://spark.apache.org/mllib/

(and shown to overcome current implicit techniques).

## B. Contributions

The major contributions of this paper are three-fold.
1) We introduce the notion of *consumed item packs* (CIPs) to extract relevant implicit information from consumption history logs of users. This notion is shown to capture the latest algorithm using implicitly consumed sequences (FISM [33]), and thus to be a general framework for reasoning about recommendation based on sequences of consumed items.
2) We propose novel algorithms using CIPs. We show first how to use CIPs to develop two memory-based techniques: a *user-based* and an *item-based* collaborative filtering algorithms, and then one model-based technique: a *neural word embedding*-based algorithm. To address scalability, these three algorithms are *incremental*: they enable to incorporate fresh items consumed recently by users, in order to update the recommendations in an efficient manner.
3) As the practical implementation and performance of the proposed algorithms are crucial to this paper, we precisely detail them in this paper. We then report on a thorough rigorous experimental evaluation in a large-scale distributed computing framework (Spark [2]), both in terms of recommendation quality and scalability. Quality results for two of our three novel algorithms is shown to always exceed state-of-the-art approaches on tested datasets.

## C. Roadmap

The rest of the paper is organized as follows. We present our notion of *consumed item packs* in § II. We present three novel algorithms in § III: the user-based variant of our recommender scheme, the item-based one, and finally the word embedding based one. We conclude this section by demonstrating in § III-D that CIPs can also capture a state-of-the-art sequential recommender. We present the implementation details of these algorithms on Spark in § IV. We present our experimental results in § V, discuss the related work in § VI, and finally conclude the paper in § VII.

## II. CONSUMED ITEM PACKS

Our consumed item packs relate to high order relations between items enjoyed by a user. Some previous works such as HOSLIM [17], considered the consumption of items by the same user as the basis for implicit recommendation. HOSLIM places the so called *user-itemsets* (implicit feedback) in a matrix, and then computes the similarity of jointly consumed items over the whole user history (that leads to the optimal recommendation quality). High-order relations are sought in principle, but due to the tractability issue of this approach (for $m$ items and order $k$: $\mathcal{O}(m^k)$ combinations of the items are enumerated and tested for relevance), authors limit computations only to pairs of items. Very recently, Barkan et al. proposed to consider item-item relations using the model of word embeddings in their technical report [10]. Our work generalizes the notion of implicit item relations, based on consumption patterns.

## A. Communities of Item Packs

To get more intuition about the very notion of *consumed item packs*, consider the following experiment we conduct on the publicly available Movielens 1M dataset, from which we extract an undirected graph. Vertices of the graph are movies. *An edge exists between two movies if some minimal number ($M$) of users have consumed both of them in a "short" consumption interval* (here "short" means consumed within - 2 to 3 contiguous hops in the users' consumption log).[2] The choice of the optimal consumption interval for specific datasets is demonstrated empirically in § V-B.

In the graph presented in Figure 1(a), we only depict, from the original graph, movies where the edges have at least 30 transitions (*i.e.*, 30 users have consumed the two movies within the specified consumption interval, leading to the representation of 1% of the total number of edges). The edges of the graph are weighted by the number of transitions, which is then at least 30 ($M = 30$).

We then apply a community detection algorithm [13] to the resulting graph. We use *modularity* as a measure of the structure of the network. The value of the modularity [13] lies in the range [-1,1]. It is positive if the number of edges within groups exceeds the number expected on the basis of chance. For a given division of the network's vertices into some modules, modularity reflects the concentration of edges within modules compared with random distribution of links between all nodes regardless of modules. A high modularity score (0.569) indicates the presence of strong communities in the graph presented in Figure 1(a). We highlight communities which represent at least 1% of the total number of nodes in the original graph. There are 10 such communities, each ranging from 1.08% to 5.21% of the original graph nodes. The average clustering coefficient of the graph is 0.475, the one of the largest community (in purple) is 0.771, and the one of the smallest community (in dark blue) is 0.842. Thus, community clustering is significantly more important than the graph one (which supports the observed high graph modularity). Interestingly, those communities are then (densely) connected, by a *latent feature*.

It is important to notice that this latent feature cannot be reduced to the *genre* of the movies. To show this, we also plot the distribution of movie genres in the 10 (strong) communities in Figure 1(b). We first observe that each community conveys a very specific blend of genres: one community cannot be trivially reduced to a genre. Secondly, it appears that some communities are closer than others: "pink" and "orange" communities are well separated, both by hop-distance on the graph (Figure 1(a)) and by their constituent genres (Figure 1(b)). The latent feature cannot be reduced to item *launch times* either: *e.g.*, movie launch times of the smallest of the 10 clusters

---
[2]The +/- signs denote the order of consumption for the pair of movies.

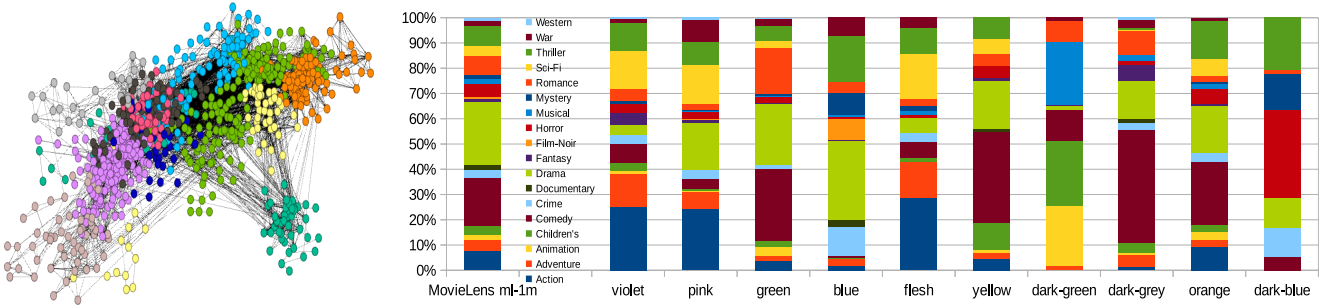

(a) **Communities of movies (Movielens).** (b) **Distribution of genres in the 10 largest communities of the movie graph. (Legend-colors on the x-axis correspond to colors of communities.**

**Fig. 1: Existence of temporal consumption habits of users in Movielens dataset.**

spread from 1931 to 1997.

We conduct a similar experiment for a product review website (Ciao [3]), setting $M = 2$ on this very sparse dataset. The resulting weighted graph, with detected item communities, also has a high modularity score of $0.61$.

In short, these experiments highlight the very existence of a non trivial latent feature, namely consumed item packs (CIPs), somehow representing the temporal consumption habits of users. Extracting this latent information from item communities and then using it for personalization services is not straightforward.

### B. Consumed Item Packs (CIPs)

To get access to this latent feature from service logs, we define the CIP data structure. CIPs are extracted from users' consumption patterns, and allow us to compute the similarity between those users (or items consumed by them). A user's profile is composed of multiple CIPs. The notion of CIP is then instantiated in three different algorithms: in a user-based algorithm (§ III-A), an item-based one (§ III-B) and a word embedding based one (§ III-C).

To make things more precise, consider a set of $m$ users $\mathcal{U} = \{u_1, u_2, ..., u_m\}$ and a set of $n$ product catalog items $\mathcal{I} = \{i_1, i_2, ..., i_n\}$. The transaction history of a user $u$, noted $\mathbf{TH}_u$, consists of a set of pairs of the form $\langle i, t_{ui} \rangle$ (where $u$ consumed an item $i$ at a time $t_{u,i}$), extracted from service logs. We denote $u$'s profile as $P_u$ which consists of the time-ordered items in $\mathbf{TH}_u$. CIPs are composed of items: each CIP $\in \mathcal{I}^*$. The order of the items in a given user's CIP represents their relative appearance in time, the leftmost symbol being the oldest one:

$\text{CIP}_u = [i_1, i_2, i_3, ..., i_k]$ such that $t_{u,i_1} < t_{u,i_2} < ... < t_{u,i_k}$. For instance, $u_1$'s CIP (CIP$_1$) is $[i_{14}, i_3, i_{20}, i_{99}, i_{53}, i_{10}, i_{25}]$, while $u_2$'s one (CIP$_2$) is $[i_{20}, i_{53}, i_4]$. Items $i_{14}$ and $i_{25}$ are respectively the first and last items that $u_1$ has consumed in CIP$_1$, while $i_{20}$ and $i_{53}$ are two items that both users have consumed. In the rest of the paper, we assume that one item occurs only once in a given CIP.[3]

A CIP then represents the items consumed by a user over a predefined period of time. Using such a data structure, one can devise a *similarity* measure $sim : \mathcal{I}^* \times \mathcal{I}^* \to \mathbb{R}^+$ between two CIPs, that captures the proximity between users (or items) as we explain in the next two sections.

In practice, CIPs are directly derived from service platform transaction logs, that are at least composed of tuples of item-id and the corresponding consumption *timestamp*[4] of that item. (It is important to note that an explicit recommender system requires tuples including, *in addition*, the rating ($r_{ui}$) that $u$ provided for item $i$.)

## III. CIP-BASED ALGORITHMS

The core claim of this paper is that the notion of CIP is general enough to capture different algorithms that rely on sequences of items. In the next three subsections, we present novel algorithms that determine CIP-based similarities and leverage sequence of items for recommendations. To illustrate the generality of CIPs, the last subsection illustrates how a previously introduced algorithm (FISM [33]) is captured by the CIP framework.

### A. CIP-U: User-based Recommender

In this subsection, we first recall the principle of a user-based CF scheme before introducing our user-based algorithm using CIPs, which we denote CIP-U. We then present how to perform incremental updates with CIP-U.

*1) User-based CF:* As depicted in Algorithm 1, a (*nearest-neighbor*) user-based CF algorithm follows two phases: *knn selection* and *recommendation*. The first phase deals with selecting $K$ most similar users (also called *neighbors*) whilst the second phase deals with recommending the most relevant items based on the profiles of these top-$K$ similar users. The similarity computation in Step 3 of Algorithm 1 employs similarity metrics like Cosine, Adjusted Cosine or Pearson Correlation [44].

*2) CIP-U Algorithm:* CIP-U is an incremental algorithm that maintains a user-user network where each user is connected to the most similar $K$ other users. CIP-U exploits users' CIPs, and accepts batches of items freshly consumed by users (*i.e.*, last logged transactions on the platform) to update this network.

---

[3]Our similarity metrics might be extended to take re-consumption into account, but it is outside the scope of this paper.

[4]The timestamp denotes the actual consumption time of the item (in the UNIX format).

**Algorithm 1** User-based CF

**Require:** $\mathcal{I}$: Set of all items; $\mathcal{U}$: Set of all users.
**Ensure:** $R_u$: Top-$N$ recommendations for user $u$.

---

***KNN selection:*** $\gamma(P_u, \mathcal{U})$ *where $P_u$ is the profile of user $u$ and $\mathcal{U}$ is the set of all users.*

---

**Require:** $\mathcal{U}, P_u$
**Ensure:** $N_u$: $K$ nearest neighbors for $u$
1: var $similarity[]$;
2: **for** $uid$ : user in $\mathcal{U}$ **do**
3:     $similarity[uid] = Sim(P_u, \mathcal{U}[uid].getProfile())$;
4: $N_u = List(sort(similarity))$;    ▷ List of sorted users
5: **return** $N_u[:K]$: $K$ users with the highest similarity;

---

***Recommendation:*** $\alpha(\mathcal{U}, P_u)$ *where $P_u$ is the profile of target user $u$.*

---

**Require:** $\mathcal{U}, N_u, P_u$
**Ensure:** $R_u$: Top-$N$ recommendations for $u$
6: var $popularity[]$;
7: **for** $uid$ : user in $N_u$ **do**
8:     **for** $iid$ : item in $\mathcal{U}[uid].getProfile()$ **do**
9:         **if** $iid \notin P_u$ **then**
10:             $popularity[iid]++$;
11: $R_u = List(sort(popularity))$;    ▷ List of sorted items
12: **return:** $R_u[:N]$: $N$ most popular items;

---

$P_u^l$ denotes the profile of a user $u$ till the $l^{th}$ update of her consumed items while $\text{CIP}_u^{l+1}$ denotes the batch of new items consumed by her since the last batch update. Assuming $P_u^l = i_1 i_2 ... i_k$ and $\text{CIP}_u^{l+1} = i_{k+1} i_{k+2} ... i_n$, we can denote the profile of a user $u$ after the $(l+1)^{th}$ iteration as $P_u^{l+1} = P_u^l \cup \text{CIP}_u^{l+1}$. Note that $\cup$ is an order preserving union here.

Before we provide the similarity measure to compare users, we introduce some preliminary definitions. We first introduce the notion of *hammock distance* between a pair of items in the profile of a given user $u$.

*DEFINITION 1 (*HAMMOCK DISTANCE*):* The hammock distance between a pair of items $(i, j)$ in $P_u$, denoted by $\mathcal{H}_u(i, j)$, is the number of hops between them.

For instance, in $P_u = [i_{14}, i_3, i_{20}, i_{99}, i_{53}, i_{10}, i_{25}]$, $\mathcal{H}_u(i_{14}, i_{99}) = 3$.

Based on the hammock distance, we define a *hammock pair* ($\mathcal{HP}$) between two users, as a pair of items that both users have in common.

*DEFINITION 2 (*HAMMOCK PAIRS*):* Given two users $u$ and $v$, their hammock pairs $\mathcal{HP}_{u,v}$ are the set of distinct item pairs both present in $P_u$ and in $P_v$, under the constraint that the number of hops between the item pairs is at most $\delta_H$.
$\mathcal{HP}_{u,v} = \{(i,j) \mid \mathcal{H}_u(i,j) \leq \delta_H \wedge \mathcal{H}_v(i,j) \leq \delta_H \wedge i \neq j\}$
Hyper-parameter $\delta_H$ denotes the *hammock threshold* and serves the purpose of tuning the CIP-based latent feature considered between related items.

Let $[\ ]$ denote the Iverson bracket:
$$[P] = \begin{cases} 1 & \text{if } P \text{ is True} \\ 0 & \text{otherwise}. \end{cases}$$

Finally, from hammock pairs, we derive the similarity of two users with regards to their CIPs as follows.

*DEFINITION 3 (*SIMILARITY MEASURE FOR USER-BASED CIP*):* The similarity between two users $u$ and $v$ is defined as a function of the cardinality of the set of hammock pairs between them:
$$sim_{\text{CIP-U}}(u,v) = 1 - (1 - [P_u = P_v]) \cdot e^{-|\mathcal{HP}_{u,v}|} \quad (1)$$
We obtain $sim_{\text{CIP-U}} \in [0, 1]$, with the boundary conditions, $sim_{\text{CIP-U}} = 0$ if the two users have no pair in common ($|\mathcal{HP}_{u,v}| = 0$ and $[P_u = P_v] = 0$), while $sim_{\text{CIP-U}} = 1$ if their CIPs are identical ($[P_u = P_v] = 1$).

*3) Incremental updates:* CIP-U enables incremental updates, in order to conveniently reflect the latest users' consumption in recommendations without requiring a prohibitive computation time. CIP-U processes batches of events (consumed items) at regular intervals and updates the similarity measure for pairs of users. $C_{u,v}$ denotes the set of items common in the profiles of two users $u$ and $v$. More precisely, after the $l^{th}$ iteration, we obtain:
$$C_{u,v}^l = P_u^l \cap P_v^l$$
Then, at the $(l+1)^{th}$ iteration, we get:
$C_{u,v}^{l+1} = P_u^{l+1} \cap P_v^{l+1}$
$= (P_u^l \cup \text{CIP}_u^{l+1}) \cap (P_v^l \cup \text{CIP}_v^{l+1})$
$= (P_u^l \cap P_v^l) \cup (P_u^l \cap \text{CIP}_v^{l+1}) \cup (P_v^l \cap \text{CIP}_u^{l+1})$
$\quad \cup (\text{CIP}_u^{l+1} \cap \text{CIP}_v^{l+1})$
$= C_{u,v}^l \cup \Delta C_{u,v}^{l+1}$
where $\Delta C_{u,v}^{l+1} = (P_u^l \cap \text{CIP}_v^{l+1}) \cup (P_v^l \cap \text{CIP}_u^{l+1}) \cup (\text{CIP}_u^{l+1} \cap \text{CIP}_v^{l+1})$. Note that the time complexity of this step is $O((|P_u^l| + |\text{CIP}_v^{l+1}|) + (|P_v^l| + |\text{CIP}_u^{l+1}|))$, where $|\text{CIP}_u^{l+1}|$, $|\text{CIP}_v^{l+1}|$ are bounded by the number of events after which the batch update will take place, say $Q$. Hence, the time complexity is $O(n + Q) = O(n)$, where $n$ denotes the total number of items, and when $Q << n$ (as expected in a system built for incremental computation).

We next incrementally compute the new hammock pairs. $\Delta \mathcal{HP}_{u,v}$ denotes the set of new hammock pairs for users $u$ and $v$. Computation is performed as follows:
$\Delta \mathcal{HP}_{u,v} = \{(i,j) \mid (i \in C_{u,v}^l, j \in \Delta C_{u,v}^{l+1}) \wedge (i \in \Delta C_{u,v}^{l+1},$
$\quad j \in \Delta C_{u,v}^{l+1}) \wedge \mathcal{H}_u(i,j) \leq \delta_H \wedge \mathcal{H}_v(i,j) \leq \delta_H\}$
The time complexity of this step is $O(|C_{u,v}^l| \cdot |\Delta C_{u,v}^{l+1}|)$, where $|\Delta C_{u,v}^{l+1}|$ is bounded by the number of events after which the batch update takes place ($Q$). Hence, the time complexity is also of $O(n \cdot Q) = O(n)$.

Finally, the similarities are computed leveraging the cardinality of the recently computed incremental hammock pairs. More precisely, we compute the updated similarity on-the-fly between a pair of users $u$ and $v$ after the $(l+1)^{th}$ iteration as follows:
$$sim_{u,v}^{l+1} = 1 - (1 - [P_u^{l+1} = P_v^{l+1}]) \cdot e^{-|\mathcal{HP}_{u,v}^l + \Delta \mathcal{HP}_{u,v}|}$$
Hence, the similarity between one user and all $m$ others is computed with a $O(nm)$ time complexity.[5] In CIP-U, we retain only a small number ($K$) of similar users. For each user $u$, we retain the $K$ most similar users, where $K \ll m$, and record these user-ids along with their similarities with $u$. We term $K$ as the *model size*. Selecting the top-$K$ similar users for collaborative filtering based on their similarity requires sorting, which induces an additional complexity of $O(m \log m)$. Hence, the total time complexity is $O(nm) + O(m \log m) = O(nm)$ (since $n \gg \log m$). Note that classical explicit collaborative filtering algorithms like user-based or item-based ones also have same time complexity for periodically updating their recommendation models. We can reduce the time complexity for the top-$K$ neighbors update further to $O(n)$ by using biased sampling and iteratively updating the neighbors [14].

---

[5] Our time complexity analysis concerns the training phase of the recommender as this phase requires more computational effort.

### B. CIP-I: Item-based Recommender

In this subsection, we first recall the principle of an item-based CF scheme and then we introduce our item-based algorithm using CIPs, which we denote as CIP-I. We then present how to perform incremental updates with CIP-I.

*1) Item-based CF:* Algorithm 2 depicts the two phases of an item-based CF. The first phase computes the top-$K$ similar items corresponding to each item for generating the item-item network where each item is connected to the most similar $K$ other items. The second phase uses this item-item network to recommend the most relevant items to the target user.

---

**Algorithm 2** Item-based CF

**Require:** $\mathcal{I}$: Set of all items; $\mathcal{U}$: Set of all users.
**Ensure:** $R_u$: Top-$N$ recommendations for $u$.

---

*KNN selection:* $\gamma(Q_i, \mathcal{I})$ *where $Q_i$ is the profile of item $i$ and $\mathcal{I}$ is the set of all items.*

---

**Require:** $\mathcal{I}$, $Q_i$
**Ensure:** $N_i$: $K$ nearest neighbors for $i$
1: var $similarity[\,]$;
2: **for** $iid$ : item in $\mathcal{I}$ **do**
3: $\quad similarity[iid] = Sim(Q_i, \mathcal{I}[iid].getProfile())$;
4: $N_i = List(sort(similarity))$; ▷ List of sorted items
5: **return:** $N_i[:K]$: $K$ most similar items;

---

*Recommendation:* $\alpha(P_u)$ *where $P_u$ is the profile of target user $u$.*

---

**Require:** $P_u$
**Ensure:** $R_u$: Top-$N$ recommendations for $u$
6: var $popularity[\,]$;
7: **for** $rid$ : item in $P_u$ **do**
8: $\quad$ **for** $iid$: item in $N_{rid}$ **do**
9: $\quad\quad$ **if** $iid \notin P_u$ **then**
10: $\quad\quad\quad popularity[iid] + +$;
11: $R_u = List(sort(popularity))$; ▷ List of sorted items
12: **return:** $R_u[:N]$: $N$ most popular items;

---

*2) CIP-I Algorithm:* CIP-I is also an incremental algorithm that processes user consumption events in CIPs, to update its item-item network.

Similar to CIP-U, we also leverage the notion of user *profiles*: a profile of a user $u$ is noted $P_u$, and is composed of one or more disjoint CIPs. We use multiple CIPs in a user profile to model her consumption pattern. CIPs are separated based on the timestamps associated with the consumed items: two consecutive CIPs are disjoint if the former's last and latter's first items are separated in time by a given interval (noted $\delta$).

*DEFINITION 4 (CIP PARTITIONS IN A USER PROFILE):* Let $i_k$ and $i_{k+1}$ denote two consecutive consumption events of a user $u$, with consumption timestamps $t_{u,i_k}$ and $t_{u,i_{k+1}}$, such that $t_{u,i_k} \leq t_{u,i_{k+1}}$. Given $i_k$ belongs to $\text{CIP}_u^l$, item $i_{k+1}$ is added to $\text{CIP}_u^l$ if $t_{u,i_{k+1}} \leq t_{u,i_k} + \delta$. Otherwise $i_{k+1}$ is added as the first element in a new $\text{CIP}_u^{l+1}$.

These CIPs are defined as $\delta$-distant. The rationale behind the creation of user profiles composed of CIPs is that each CIP is intended to capture the semantic taste of a user within a consistent consumption period.

With $i <_{\text{CIP}} j$ denoting the prior occurrence of $i$ before $j$ in a given CIP, and the inverse hammock distance ($\epsilon_u(i,j)$) being a penalty function for distant items in a $\text{CIP}_u$ (*e.g.*, $\epsilon_u(i,j) = \frac{1}{\mathcal{H}_u(i,j)}$), we express a similarity measure for items, based on those partitioned user profiles, as follows.

*DEFINITION 5 (SIMILARITY MEASURE FOR ITEM-BASED CIP):* Given a pair of items $(i,j)$, their similarity ($sim_{\text{CIP-I}}(i,j) = s$) is:

$$s = \frac{\sum_u \sum_{l=1}^{|l|_u} [(i,j) \in \text{CIP}_u^l \land i <_{\text{CIP}} j](1 + \epsilon_u(i,j))}{2 \cdot \max\{\sum_u \sum_{l=1}^{|l|_u} [i \in \text{CIP}_u^l], \sum_u \sum_{l=1}^{|l|_u} [j \in \text{CIP}_u^l]\}}$$
$$= \frac{score_{\text{CIP-I}}(i,j)}{2 \cdot max\{cardV(i), cardV(j)\}}$$
(2)

where $|l|_u$ denotes the number of CIPs in the profile of user $u$ and $[\,]$ denotes the Iverson bracket.

This reflects the number of close and ordered co-occurrences of items $i$ and $j$ over the total number of occurrences of both items independently: $sim_{\text{CIP-I}}(i,j) = 1$ if each appearance of $i$ is immediately followed by $j$ in the current CIP. Contrarily, $sim_{\text{CIP-I}}(i,j) = 0$ if there is no co-occurrence of those items in any CIP. Furthermore, we denote the numerator term as $score_{\text{CIP-I}}(i,j)$ and the denominator term as a function of $cardV(i)$ and $cardV(j)$ sub-terms for Equation 2 where $cardV(i) = \sum_u \sum_{l=1}^{|l|_u} [i \in \text{CIP}_u^l]$. As shown in Algorithm 3, we can update $score_{\text{CIP-I}}(i,j)$ and $cardV(i)$ terms incrementally. Finally, we can compute the similarity on-the-fly leveraging $score_{\text{CIP-I}}(i,j)$ and $cardV(i)$ terms.

*3) Incremental updates:* CIP-I processes users' recent CIPs scanned from users' consumption logs. Score values ($score_{\text{CIP-I}}$) are updated as shown in Algorithm 3. We require an item-item matrix to maintain the *score* values, as well as an $n$-dimensional vector that maintains the current occurrence number of each item.

After the update of the *score* values, the algorithm terminates by updating a data structure containing the top-$K$ closest items for each given item, leveraging the *score* matrix and the cardinality terms for computing similarities on-the-fly.

The complexity of Algorithm 3 depends on the maximum tolerated size of incoming CIPs. As one expects an incremental algorithm to receive relatively small inputs as compared to the total dataset size, the final complexity is compatible with online computation: *e.g.*, if the largest CIP allowed has cardinality $|\text{CIP}| = O(\log n)$, then run-time complexity is

**Algorithm 3** *Incremental Updates for Item Pairs.*

**Require:** $CIP_u$ ▷ last $\delta$-distant CIP received for user $u$
1: $score_{\text{CIP-I}}[\ ][\ ]$ ▷ item-item *score* matrix, intialized to 0
2: $cardV$ ▷$n$-dim. vector of appearance cardinality of items
3: **for** item $i$ in $CIP_u$ **do**
4:     $cardV(i) = cardV(i) + 1$
5:     **for** item $j$ in $CIP_u$ **do**
6:         **if** $i \neq j$ **then**
7:             $\epsilon(i,j) = \epsilon(j,i) = \frac{1}{\mathcal{H}_u(i,j)}$
8:         **if** $i <_{\text{CIP}} j$ **then**
9:             $score_{\text{CIP-I}}[i][j]+=(1+\epsilon(i,j))$
10:         **else**
11:             $score_{\text{CIP-I}}[j][i]+=(1+\epsilon(j,i))$

poly-logarithmic.

### C. DEEPCIP: Embedding-Based Recommender

In this subsection, we present an approach based on machine learning, inspired by WORD2VEC [10], [40]. This approach relies on word embedding, transposed to items. We specifically adapt this concept to our CIP data structure. We name this CIP-based approach DEEPCIP.

*1) WORD2VEC Embeddings:* Neural word embeddings, introduced in [12], [40], are learned vector representations for each word from a text corpus. These neural word embeddings are useful for predicting the surrounding words in a sentence. A common approach is to use a multi-layer Skip-gram model with negative sampling. The objective function minimizes the distance of each word with its surrounding words within a sentence while maximizing the distances to randomly chosen set of words (*negative samples*) that are not expected to be close to the target. This is an objective quite similar to ours as it enables to compute proximity between items in the same CIP. This approach computes similarity between two words as the dot product of their word embeddings.

*2) DEEPCIP Algorithm:* We now describe how the WORD2VEC concept is adapted to CIPs, for they allow scalable and fresh item incorporation in the model. We feed a skip-gram model with item-pairs in CIPs where each CIP is as usual an ordered set of items (similar to the instantiation in CIP-I). More precisely, CIPs are $\delta$-distant as instantiated in § III-B. DEEPCIP trains the neural network with pairs of items at a distance less than a given *window size* within a CIP. This window size corresponds to the notion of hammock distance (defined in § III-A) where the distance hyper-parameter $\delta_H$ is defined by the *window size*. More formally, given a sequence of $T$ training items' vectors $i_1, i_2, i_3, ..., i_T$, and a maximum hammock distance of $k$, the objective of the DEEPCIP model is to maximize the average log probability

$$\frac{1}{T} \sum_{t=k}^{T-k} \log P(i_t | i_{t-k}, ...., i_{t-1}, i_{t+1}, ...., i_{t+k}) \quad (3)$$

The Skip-gram model is employed to solve the optimization objective 3 where the weights of the model are learned using backpropagation and stochastic gradient descent (SGD). SGD is inherently synchronous as there is a dependence between the update from one iteration and the computation in the next iteration. Each iteration must potentially wait for the update from the previous iteration to complete. This approach does not allow the distribution of computations on parallel resources which leads to a scalability issue. To circumvent this scalability issue, we implement DEEPCIP using asynchronous stochastic gradient descent (DOWNPOUR-SGD [22]). DOWNPOUR-SGD enables distributed training for the skip-gram model on multiple machines by leveraging asynchronous updates from them. We use a publicly-available deep learning framework [4] which implements DOWNPOUR-SGD in a distributed setting. More precisely, DEEPCIP trains the model using DOWNPOUR-SGD on the recent CIPs thereby updating the model incrementally.

DEEPCIP uses a *most_similar* functionality to select items to recommend to a user, using as input recently consumed items (current CIP). We compute a CIP vector using the items in the given CIP and then use this vector to find most similar other items. More precisely, the *most_similar* method uses the cosine similarity between a simple mean of the projection weight vectors of the recently consumed items (i.e., items in a user's most recent CIP) and the vectors for each item in the database.

*3) Incremental updates:* Online machine learning is performed to update a model when data becomes available. The DEEPCIP model training is performed in an online manner [23] where the model is updated using the recent CIPs. Online machine learning is crucial in recommendation as it is necessary for the algorithm to dynamically adapt to new temporal patterns [16], [26], [29] in the data. Hence, the complexity of the model update is dependent on the number of new CIPs received along with the hyper-parameters for the learning algorithm (primarily, skip-gram model parameters, dimensionality of item vectors, number of training iterations, hammock distance).

### D. The FISM algorithm under CIPs

We now demonstrate that the CIP framework can incorporate the state-of-art sequence-based algorithm FISM [33], in order to illustrate the generality of the CIP notion.

In FISM, the item-item similarity is computed as a product of two low-ranked matrices $\mathbf{P} \in \mathcal{R}^{m \times k}$ and $\mathbf{Q} \in \mathcal{R}^{m \times k}$ where $k << m$. More precisely, the item-item similarity between any two items is defined as $sim(i,j) = \mathbf{p}_j \mathbf{q}_i^T$ where $\mathbf{p}_j \in \mathbf{P}$ and $\mathbf{q}_i \in \mathbf{Q}$. Finally, the recommendation score for a user $u$ on an unrated item $i$ (denoted by $\bar{r}_{ui}$) is calculated as an aggregation of the items that have been rated by $u$.

$$\bar{r}_{ui} = b_u + b_i + (n_u^+)^{-\alpha} \sum_{j \in \mathcal{R}_u^+} \mathbf{p}_j \mathbf{q}_i^T \quad (4)$$

where $\mathcal{R}_u^+$ is the set of items rated by user $u$ (note that FISM do not leverage ratings, but only the fact that a rated item has been consumed by definition), $b_u$ and $b_i$ are the user and item biases, $\mathbf{p}_j$ and $\mathbf{q}_i$ are the learned item latent factors, $n_u^+$ is the number of items rated by $u$, and $\alpha$ is a user specified parameter between 0 and 1. Moreover, term $(n_u^+)^{-\alpha}$ in Equation 4 is used to control the degree of agreement between the items rated by the user with respect to their similarity to the item whose rating is being estimated (*i.e.*, item i).

We now present how Equation 4 is adapted to fit into the CIP notion. For a user $u$, we denote her profile ($P_u$) consists of $|l|_u$ different CIPs (similar to the notations introduced for Equation 2). Equation 4 can be rewritten using CIPs as follows:

$$\bar{r}_{ui} = b_u + b_i + (|\cup_{k=1}^{|l|_u} \text{CIP}_u^k|)^{-\alpha} \sum_{k=1}^{|l|_u} \sum_{j \in \text{CIP}_u^k} \mathbf{p}_j \mathbf{q}_i^T, \quad (5)$$

where |·| denotes the cardinality and we substitute consumed items by CIP structures; this last transformation shows that indeed CIPs incorporates the FISM definition of item sequences. We also note that due to the CIPs, the terms in Equation 5 could be incrementally updated, similar to CIP-U and CIP-I, by incorporating the latest CIP for the given user.

## IV. IMPLEMENTATION

We provide here some implementation details of our CIP-U, CIP-I and DEEPCIP algorithms.

### A. Spark Data Structures

We consider Apache Spark [2] as our framework for recommendation computations. Spark is a cluster computing framework for large-scale data processing. It is built on top of the Hadoop Distributed File System (HDFS) and provides several core abstractions, namely Resilient Distributed Datasets (RDDs), parallel operations and shared variables.

An RDD is a fault-tolerant abstraction that enables users to explicitly persist intermediate results in memory and control their partitioning to optimize data placement. It is a read-only collection of objects partitioned across a set of machines and can be rebuilt if a partition is lost. In a Spark program, data is first read into an RDD object. This RDD object can be altered into other RDD objects by using *transformation operations* like `map`, `filter`, and `collect`. Spark also enables the use of shared variables, such as *broadcast* and *accumulator*, for accessing or updating shared data across worker nodes.

### B. Tailored Data Structures for CIPs

We now mention briefly the RDDs leveraged in the memory-based approaches (CIP-U and CIP-I).

*1) RDDs for CIP-U:* For CIP-U, we store the collected information into three primary RDDs as follows. USERSRDD stores the information about the user profiles. USERSIMRDD stores the hammock pairs between all pairs of users. The pairwise user similarities are computed using a transformation operation over this RDD. USERTOPKRDD stores the $K$ most similar users.

During each update step in CIP-U, after $Q$ consumption events, the new events are stored into a DELTAPROFILES RDD which is broadcast to all the executors using the *broadcast* abstraction of Spark. Then, the hammock pairs between users are updated (in USERSIMRDD) and consequently transformed to pairwise user similarities using Equation 1. Finally, CIP-U updates the the top-$K$ neighbors (USERTOPKRDD) based on the updated similarities.

*2) RDDs for CIP-I:* For CIP-I, we store the collected information into two primary RDDs as follows. ITEMSIMRDD stores *score* values between items. The pairwise item similarities are computed using a transformation operation over this RDD. ITEMTOPKRDD stores the $K$ most similar items for each item based on the updated similarities.

During each update step in CIP-I, the item scores are updated incorporating the received CIP using Algorithm 3 in the ITEMSIMRDD, and consequently the pairwise item similarities are also revised using Equation 2. CIP-I computes the top-$K$ similar items and updates the ITEMTOPKRDD at regular intervals.

*3) RDDs for DEEPCIP:* We implement the DEEPCIP using the DeepDist deep learning framework [4] which accelerates model training by providing asynchronous stochastic gradient descent (DOWNPOUR-SGD) for data stored on Spark.

DEEPCIP implements a standard master-workers parameter server model [22]. On the master node, the CIPSRDD stores the recent CIPs aggregated from the user transaction logs preserving the consumption order. DEEPCIP trains on this RDD using the DOWNPOUR-SGD. The skip-gram model is stored on the master node and the worker nodes fetch the model before processing each partition, and send the gradient updates to the master node. The master node performs the stochastic gradient descent asynchronously using the updates sent by the worker nodes. Finally, DEEPCIP predicts the most similar items to a given user, based on her most recent CIP.

## V. EVALUATION

In this section, we report on the evaluation of the CIP-based algorithms, using real-world datasets. We measure their recommendation quality along with their scalability for processing incoming batches of consumption events.

### A. Experimental Setup

We now provide the details regarding our experimental setup mainly in terms of the deployment platform, datasets, evaluation metrics and evaluation scheme. Lastly, we also present a brief overview for each of the competitors which we compare our CIP-based algorithms with.

*1) Platform:* For our experiments, we use two deployment modes of the Spark large-scale processing framework [2].

*Standalone deployment.* We launch a Spark Standalone cluster on a highperf server (Dell Poweredge R930) with 4 Processors Intel(R) Xeon(R) E7-4830 v3 (12 cores, 30MB cache, hyper-threading enabled) and 512 GB of RAM. We use this cluster to evaluate the effect of the number of partitions for the RDD on scalability. For the standalone deployment, we use 19 executors each with 5 cores since we have a total of 96 cores in this cluster.[6]

*YARN deployment.* We use the Grid5000 testbed to launch a Spark cluster consisting of 20 machines on Hadoop YARN. Each machine is an Intel Xeon CPU E5520@ 2.26GHz. For the YARN deployment, we set the number of executors equal to the number of machines in the cluster.

---
[6]We use this deployment for running long duration experiments, due to reservation limitations on the Grid5000 cluster.

| Datasets | #Users, #Items | #Training, #Validation, #Test | Density |
|---|---|---|---|
| ML-100K | 943, 1682 | 75000, 5000, 20000 | 6.31% |
| ML-1M | 6040, 3952 | 970209, 10000, 20000 | 4.19% |
| Ciao | 489, 12679 | 19396, 1000, 2000 | 0.36% |

**TABLE I: Details of the datasets used in our experiments.**

*2) Datasets:* We use real-world traces from a movie recommendation website: Movielens (ML-100K, ML-1M) [5] as well as a product review website: Ciao [3]. Those traces contain users' ratings for movies they enjoyed. We compare the performance of our implicit CIP based models to the one of a widespread explicit (rating-based) collaborative filtering. In these datasets, each user rated at least 20 movies. The ratings vary from 1 to 5 with an increment of 1 between the possible ratings. Note that the ratings are only used for the explicit (rating-based) recommender. Table I provides further details about these datasets along with their *densities*. The density of a dataset denotes the fraction of actual user-item (implicit or explicit) interactions present in the dataset compared to all the possible interactions.

*3) Metrics:* Our evaluation consists of two complementary metrics: (a) the recommendation quality as perceived by the users in terms of *precision*, and (b) the scalability in terms of *speedup* over the computations achieved by increasing the number of machines in a cluster.

(a) *Quality.* We evaluate the recommendation quality in terms of the *precision* which is a classification accuracy metric used conventionally to evaluate top-$N$ recommenders [19]. Precision denotes the fraction of recommended items which were indeed relevant to the target user.

(b) *Scalability.* We evaluate the scalability in terms of the *speedup* over the computations which compares the time required for parallel execution with $p$ processors ($T_p$) with respect to the time required for sequential execution ($T_1$). Amdahl's law [35] models the performance of speedup ($S_p$) as: $S_p = T_1/T_p$.

*4) Evaluation scheme:* The dataset is sorted based on the unix timestamps associated with the rating events. Then, the sorted dataset is replayed to simulate the actual temporal behavior of users. We measure the recommendation quality as follows: we divide the sorted dataset into a *training set*, a *validation set* and a *test set*. The training set is used to train our CIP based models whereas the validation set is used to tune the hyper-parameters of the models. For each event in the test set (or rating when applied to explicit recommenders), a set of top recommendations is selected as the *recommendation set* with size denoted as $N$. Note that we recommend the most popular items for new users (cold-start). Table I shows the partition between training, validation and test sets along with the details of the datasets. For scalability tests, we measure the speedup for the incremental updates in the training phase while increasing the number of machines (or partitions) in the cluster.

*5) Hyper-parameters:* We tune the core hyper-parameters for CIP-U, CIP-I and DEEPCIP. For CIP-U, we have the *hammock threshold* ($\delta_H$) whereas for the CIP-I, we have the distance ($\delta$) to separate $\delta$-distant CIPs in a user's profile. For DEEPCIP, we have the distance ($\delta$), similar to CIP-I, as well as the window size ($W$) which denotes the maximum hop allowed for learning the item vectors within a CIP. These hyper-parameters essentially determine the optimal size of the consumption interval for achieving the best recommendation quality.

*6) Competitors:* We compare the recommendation quality of our three algorithms with also three competitors: a *matrix factorization* based technique (using explicit ratings) [34], a popular time-based recommender (without using any explicit ratings) [36], and the state-of-the art approach mixing both implicit and explicit information [39].

**Matrix factorization.** Matrix factorization techniques map both users and items to a joint latent factor space of dimensionality $f$, such that ratings are modeled as inner products in that space. We use a publicly available library (Python-recsys [6]) for empirical evaluations. Python-recsys is a widely used recommender framework for SVD-based approaches [45], [49].

**Implicit time-based recommender.** We compare with a popular time-based recommender designed to provide recommendations without the need for explicit feedback [36]. They construct pseudo ratings from the collected implicit feedback based on temporal information - *user purchase-time* and *item launch-time* - in order to improve recommendation accuracy. They use two rating functions: $W_3$ (coarse function with three launch-time groups and three purchase-time groups) and $W_5$ (fine-grained function with five launch-time groups and five purchase-time groups) where the later performs slightly better. Hence, we choose $W_5$ rating function for our empirical comparison and we denote this system as $TB - W_5$ in our evaluation.

**Markov chain-based recommender.** We compare with a recent recommender which combines matrix factorization and markov chains [39], [43] to model personalized sequential behavior. We use a publicly available library [7] for our empirical evaluation. We denote this system as MCREC in our evaluation. We note that we do not compare with FISM [33], as it is empirically shown to be outperformed by the markov-chain based ones.

### B. Quality

We now provide the results of the recommendation quality evaluation for our three algorithms.

*1) Hammock threshold in CIP-U:* Figure 2 demonstrates the impact of the hammock threshold ($\delta_H$) as well as the model size ($K$) on the recommendation quality in terms of precision. We observe that the quality improves as the CIP-U model size increases, and after an optimal model size, the quality starts decreasing. This behavior is due to the fact that a significantly higher model size includes neighbors who are not so similar thereby impacting quality negatively. Furthermore, we also observe that as the hammock threshold increases the recommendation quality, in terms of precision, it improves until a peak point, before it starts decreasing. This decrease in quality is due to the fact that increasing the hammock

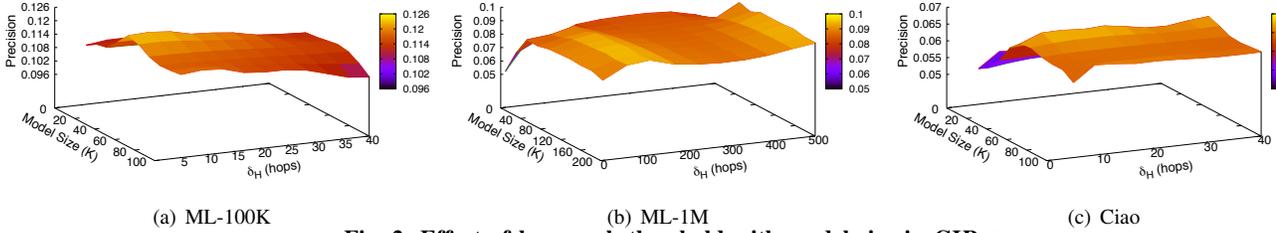

(a) ML-100K  (b) ML-1M  (c) Ciao
**Fig. 2: Effect of hammock threshold with model size in CIP-U.**

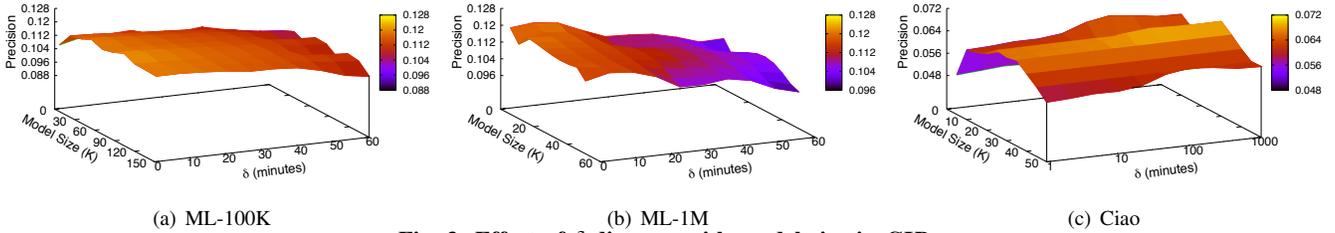

(a) ML-100K  (b) ML-1M  (c) Ciao
**Fig. 3: Effect of $\delta$-distance with model size in CIP-I.**

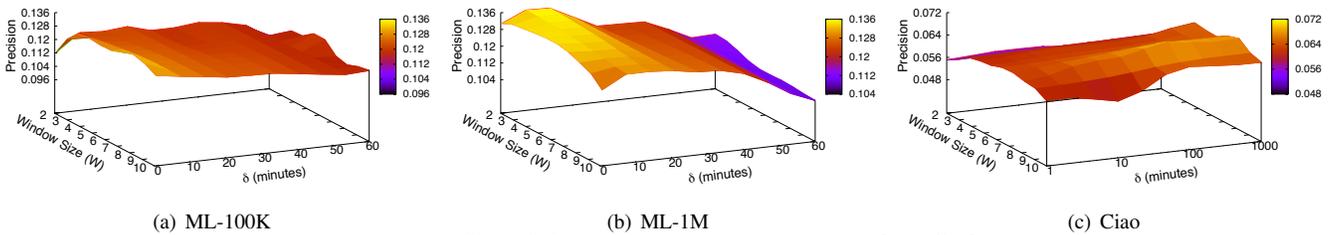

(a) ML-100K  (b) ML-1M  (c) Ciao
**Fig. 4: Effect of $\delta$-distance with window size in DEEPCIP.**

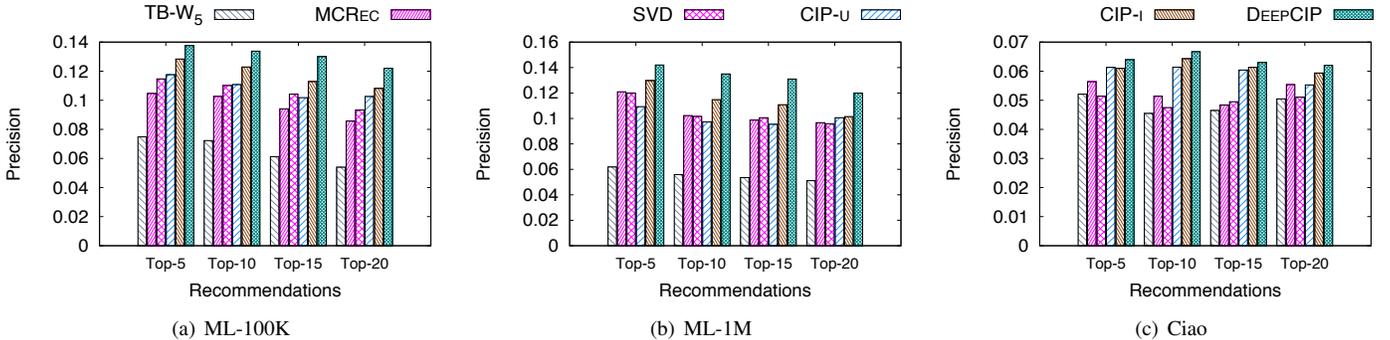

(a) ML-100K  (b) ML-1M  (c) Ciao
**Fig. 5: Recommendation quality of CIP-based algorithms versus competitors.**

threshold would result in more hammock pairs between two user profiles which have larger hammock distances and hence are less relevant due to the substantial gap in their consumption instances. Based on our observations, as shown in Figure 2, we set $\delta_H = 10$ for ML-100K, $\delta_H = 30$ for ML-1M and $\delta_H = 10$ for Ciao to attain the best possible quality. We set the model size to $K = 50$ for further experiments.

*2) $\delta$-distance in CIP-I:* The distance hyper-parameter varies according to datasets, depending on the nature of consumption interaction. For example, for rating/review services (Movielens, Ciao, Imdb), browsing/purchasing services (Amazon, eBay), it can be based on the interval between two consumption events whereas for entertainment services like TiVo or Netflix, it can be based on per log-in session. Figure 3 demonstrates the impact of the distance ($\delta$) as well as the model size ($K$) on the recommendation quality in terms of precision. We observe a similar behavior of the model size on the quality as in CIP-U. We also observe that as the distance increases the recommendation quality improves until a point and then it decreases. The behavior in recommendation quality is due to the fact that as the distance increases, the CIPs include less temporally relevant items. Based on our observations, as shown in Figure 3, we set $\delta = 1$ minute for ML-100K, $\delta = 1$ minute for ML-1M and $\delta = 100$ minutes for Ciao to attain the best possible quality. The relatively smaller value for $\delta$ for Movielens is due to the fact that Movielens records *rating sessions* of its users which consist of multiple movies rated within a short time interval (*i.e.*, users are rating batches of movies). We set the model size to $K = 30$ for further experiments.

*3) $\delta$-distance in DEEPCIP:* Similar to CIP-I, we vary the distance as well as the window size hyper-parameters. Figure 4

demonstrates the impact of the distance ($\delta$) as well as the window size ($W$) hyper-parameters on the recommendation quality in terms of precision. We observe that the quality decreases if the window size exceeds some threshold. This decrease can be attributed to the fact that the skip-gram model learns item vectors based on item-pairs which are not so much related (temporally) and thereby generates noisy item vectors. Based on our observations, as shown in Figure 4, we set $\delta = 1$ minute for ML-100K, $\delta = 1$ minute for ML-1M and $\delta = 100$ minutes for Ciao to attain the best possible quality. We set the window size ($W$) to 5 for all three datasets.

Additionally, we set the recommendation set size ($N$) to 10 for all further experiments. A value of $N = 10$ denotes that for every click the user will be recommended the ten most relevant items. For the matrix factorization techniques, we set the number of features used to 50, for Movielens as well as Ciao, as we observed that the precision saturates at 50 features on both the datasets.

*4) Comparison with competitors:* Once we obtain the optimal setting of the hyper-parameters for our CIP based models, we compare them with the competitors namely: the matrix factorization based technique (SVD), the markov-chain based technique (MCREC) and the time-based approach (TB-$W_5$). We compare the recommendation quality in terms of the precision ($N = 10$) on Movielens (ML-100K, ML-1M) and Ciao datasets, in Figure 5. We draw the following observations.

- Regarding our three algorithms, DEEPCIP always outperforms CIP-I, which in turn is always outperforming CIP-U (except on the Top-5 result on the Ciao dataset which is due to the relatively limited number of recommendations).
- The CIP-based algorithms outperform TB-$W_5$ on all the three datasets. For example, consider top-10 recommendations in the ML-1M dataset, CIP-U provides around $1.82\times$ improvement in the precision, CIP-I provides around $2.1\times$ improvement, and DEEPCIP provides around $2.4\times$ improvement.
- The CIP-U algorithm performs on par with MCREC as well as matrix factorization based techniques. CIP-I overcomes MCREC on all three scenarios, sometimes only by a short margin (ML-1M). However, the DEEPCIP model outperforms all other models significantly. For example, consider the top-10 recommendations in the ML-1M dataset, DEEPCIP provides $2.4\times$ improvement over TB-$W_5$, $1.29\times$ improvement over MCREC, and $1.31\times$ improvement over the matrix factorization based one. The reason behind this improvement is that DEEPCIP considers, for any given item, the *packs* of items at a distance dependent on the defined window size, whereas MCREC only considers pairs of items in the sequence of chain states (and thus has a more constrained learning process).

Note that the precision we obtain for SVD on Movielens (11% to 12%) is consistent with other standard quality evaluation benchmarks for state-of-the-art recommenders [19].

These results show the existence of the latent information contained in closely consumed items, accurately captured by the CIP structure. Note that this is intuitively consistent for DEEPCIP to perform well in this setting: the original WORD2VEC concept captures relation among words w.r.t. their proximity in a given context. With DEEPCIP, we seek to capture item proximity w.r.t. their consumption time.

*C. Scalability*

*1) Number of partitions:* Spark's RDD deals with fragmented data which enables Spark to efficiently execute computations in parallel. The level of fragmentation is a function of the number of partitions of an RDD which is crucial for the scalability performance of an application. A small number of partitions reduces the concurrency and consequently leads to under-utilization of the cluster. Furthermore, since with fewer partitions there is more data in each partition, this increases the memory pressure on the application. On the flip side, with too many partitions, the performance might degrade due to data shuffling as it takes a hit from the network overheads and disk I/Os. Hence, tuning the number of partitions is important in determining the attainable scalability of an algorithm. We thus conduce the effect of the number of partitions on scalability. We run these experiments in the Standalone mode of Spark.

Figures 6(a) and 6(b) demonstrate that scalability depends on the number of partitions which is ideally equal to the number of cores in the cluster. We observe a near-linear speedup while increasing the number of partitions for both CIP-U as well as DEEPCIP. However, the speedup is comparatively less for CIP-I due to the highly reduced time complexity of CIP-I leading to significantly less computations.

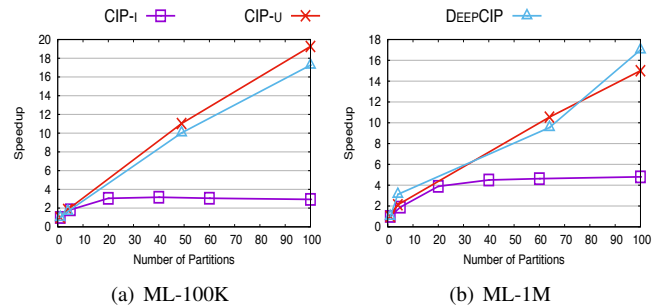

(a) ML-100K     (b) ML-1M

Fig. 6: Partition effects.

*2) Cluster size:* We now evaluate the scalability of our algorithms while increasing the cluster size from one machine to a maximum of 20 machines. Furthermore, we also compare the speedup achieved by a matrix factorization technique (ALS) implemented in the publicly available MLLIB library for Spark. Number of partitions is set to 50.

Figure 7 depicts a sublinear increase in speedup while increasing the number of machines on both the datasets. The sublinearity in the speedup is due to communication overheads in Spark with increasing number of machines. The speedup on ML-1M is higher due to more computations being required for larger datasets and higher utilization of the cluster. We observe that the speedup for CIP-I is similar for both datasets as its time complexity depends on the CIP size (Algorithm 3). DEEPCIP scales well due to the distributed asynchronous stochastic gradient descent (DOWNPOUR-SGD) for training the skip-gram model where more gradient computations could be executed asynchronously in parallel with increasing number of nodes. CIP-U and DEEPCIP scale better than ALS for both setups.

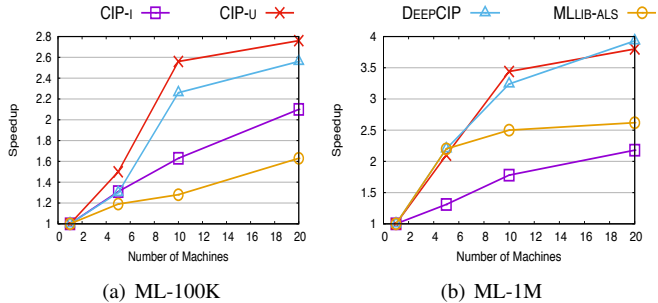

(a) ML-100K  (b) ML-1M

**Fig. 7: Cluster size effects.**

## VI. RELATED WORK

We now discuss previous work about using explicit and implicit feedback in recommenders.

### A. Explicit feedback

Tapestry [24], one of the earliest implementations of collaborative filtering, relies on the explicit opinions of people from a close-knit community such as an office working group. Since then, a lot of work has been devoted to improve the recommendation quality and even incremental updates [14], [20]. All however require explicit feedback like numerical ratings, binary like/dislike or just positive likes. Recently, Sen et al. demonstrated that different rating scales elicit different levels of cognitive load on the end users [47]. Whitenton pointed out the relation between cognitive load and consumer usability and highlighted the very fact that to achieve maximum usability, the cognitive load should be minimized [48]. In this paper, we focus on utilizing the information available in transaction logs, for it is available to arguably all services.

### B. Implicit feedback

Our CIP-based algorithms belong to the category of recommenders using implicit feedback from users [41]. HOSLIM [17] proposes to compute higher order relations between items in consumed itemsets; those relations are the ones that maximize the recommendation quality, but without notions of temporality in item consumption. The proposed algorithm is time-agnostic, and does not scale for orders superior to pairs of items. Moreover, it is not designed to efficiently incorporate freshly consumed items and suffers from computational intractability. Barkan et al. present ITEM2VEC in their technical report [10], that also uses skip-gram with negative sampling to retrieve items' relations w.r.t their context in time. Besides the fact that their implementation does not scale on multiple machines due to the use of synchronous stochastic gradient descent, their technical report is also not detailed, and evaluated only on private datasets. This makes precise evaluations w.r.t. state-of-the-art algorithms subjective. Implicit feedback has also been used for multiple other applications: this is traditionally the case in search engines, where clicks are tracked [18]. SPrank [42] leverages semantic descriptions of items, gathered in a knowledge base available on the web. Koren et al. [30] showed that implicit information, like channel switching on TV, is valuable enough to propose recommendations. Huang et al. leverage unordered co-occurrence of contextual queries in session-based query logs in a non-incremental manner for relevant term suggestion in search engines [31]. Recommenders can also use the implicit social information of their users to improve final results [38].

Interestingly enough, in the context of music recommendation, Jawaheer et al. [32] pointed out that implicit and explicit recommenders are complementary, and experimentally perform similarly. Recently, Soldo et al. leveraged users' malicious (implicit) activity logs to recommend which IP addresses to block [46]. Hence, implicit feedback based approaches could be employed over a wide range of applications.

*1) Time-based recommendation:* Within implicit based recommenders, the notion of "time" has been exploited in various ways since it is a crucial implicit information collected by all services. Some companies implement implicit recommenders, as *e.g.*, Amazon [1]; yet, we are not aware of the use of any technique even remotely close to our notion of item packs. The use of spatio-temporal proximity between users in a given place was introduced in [21]. However, such a technique requires auxiliary location-based information for detecting such user proximity, which furthermore might be a privacy concern for users (*location privacy* [11]). Baltrunas et al. presented a technique [9] very similar to CIP where a user profile is partitioned into micro-profiles (similar to CIPs in our approach). However, explicit feedback is required for each of these micro-profiles, to improve the quality of recommendations. Time window (or decay) filtering is another technique, applied to attenuate recommendation scores for items having a small likelihood to be purchased at the moment when a user might view them [25]. While such an approach uses the notion of time in transaction logs to improve recommendations, it still builds on explicit ratings for computing the basic recommendation scores. Campos et al. [15] proposed to bias recommendation according to freshness of ratings in the dataset. However, their approach still uses explicit ratings to improve recommendation quality using their time-biased strategy. Finally, Lee et al. [36] introduced a completely implicit feedback based approach that gives more weight to new items if users are sensitive to the item's launch times. We compare our algorithms to this approach in § V-B and demonstrate that our CIP-based algorithms perform better.

*2) Sequence-based recommendation:* Recently, there have been some approaches using Markov chains to model consumption sequences [43]. However, such approaches suffer from sparsity issues and the long-tailed distribution of many datasets. We compare with a Markov-chain based approach (MCREC) in § V-B and show that CIP-based approaches, updated incrementally in a distributed manner, perform on par with MCREC.

## VII. Conclusion

Since very recently, research efforts are dedicated to circumvent the absence of explicit feedback on online platforms, using individual techniques that leverage the sequential consumption of items. In an effort for a detailed and scalable proposal for generalizing such a direction, we presented two memory-based and one model-based recommendation algorithms exploiting the implicit notion of *item packs consumed by users*, while showing that our framework can also incorporate the previous state-of-the-art approach on the topic. Our novel algorithms provide a better recommendation quality than the widespread SVD-based approach [34], as well as implicit ones leveraging consumption time [36] or consumption sequences [39], [43]. This confirms the fact that item packs allow to efficiently identify similar users or items, as illustrated in § II. Importantly, for practical deployments, this key latent feature can be captured with the incremental algorithms that we presented, thus allowing to build fast services using freshly consumed items. Lastly, it would be interesting to incorporate privacy [27], [28] for users with such incremental algorithms.